\documentstyle{elsart}

\input epsf

\begin{document}
\begin{frontmatter}
\title{
Interfaces in partly compatible polymer mixtures: a Monte Carlo simulation 
approach
}
\author{Kurt Binder, Marcus M\"uller, Friederike Schmid, and Andreas Werner}
\address{
Institut f\"ur Physik, Johannes Gutenberg Universit\"at Mainz
D 55099 Mainz, Staudinger Weg 7, Germany}

\begin{abstract}
The structure of polymer coils near interfaces between 
coexisting phases of symmetrical polymer mixtures (AB) is discussed, 
as well as the structure of symmetric diblock copolymers of the same
chain length $N$ adsorbed at the interface. The problem is studied
by Monte Carlo simulations of the bond fluctuation model on the
simple cubic lattice, choosing $N=32$ and lattice linear dimensions
$L \times D \times L$ up to $512 \times 64 \times 512$, and using 
massively parallel computers (CRAY T3D).
While homopolymer coils in the strong segregation limit are oriented parallel
to the interface, the diblocks form ``dumbbells'' oriented perpendicular
to the interface. However, in the dilute case (``mushroom regime'' rather 
than ``brush regime''), the diblocks are only weakly stretched.
Distribution functions for monomers at the chain ends and in the center 
of the polymer are obtained, and
a comparison to the self consistent field theory is made.
\end{abstract}
\end{frontmatter}

\section{Introduction}
Blending of polymers is useful to obtain materials with improved properties\cite{1}.
However, most pair (A,B) of chemically different homopolymers do not mix:
any unfavorable enthalpy per monomer, multiplied by the large number of monomers in a chain
(``chain length'' $N_A,N_B$; for simplicity only `` symmetric mixtures'', $N_A=N_B=N$ are 
treated here), will exceed the entropy of mixing (which is of order $k_BT$, $T$ being the
absolute temperature)\cite{2}. Therefore 
the widely used polymer blends are not homogeneous on mesoscopic scales,
rather they are fine dispersion of one polymer in another\cite{3},i.e.\ the material is full
of AB-interfaces, and the detailed structure of the polymers in the interfacial region clearly 
have a crucial importance for the understanding the application properties of the blend.
\newline
In practice one often uses block copolymers \cite{4,5} as polymeric surfactants that control
the size of the minority droplets in the dispersion. The classical explanation\cite{1,6}
for the usefulness of the block copolymer to produce fine enough dispersion is based on the
fact that the adsorption of block copolymers at the AB interface lowers the interfacial
tension\cite{7,8}. However, this mechanism was recently questioned\cite{3,9,10}, and it
was argued the primary effect of the adsorbed block copolymer is a kinetic one, i.e.\
they prevent the small droplets to a large extent from coagulation.
\newline
For a more complete understanding one clearly needs detailed knowledge of interfacial
properties, both with and without adsorbed copolymers, including the detailed configurations
of the chains when they are confined to the interfacial region. While numerous
elegant experiments have yielded already interesting information on this
problem (e.g.\ \cite{11,12,13,14,15,16}), computer simulation\cite{17,18,19} is a tool that can provide a much more detailed
and complete picture. Although the models accessible
to simulation\cite{20} are very idealized, they are well defined and all the
quantities needed to stringently test corresponding theories such as the self consistent field (SCF)
theory\cite{8,18,19,21,22,23} can easily be estimated from such simulations. There is no problem
with fitting of phenomenological parameters to experiments. In the present paper, we focus on the configurational
properties of the chains in the interfacial region of a strongly segregated mixture, considering also adsorbed 
diblock copolymers of the same chemical nature AB, symmetric composition ($f=1/2$)\cite{4,5}, and the same
chain length as the homopolymers.

\section{Some comments on the model and the simulation method}
While simulational efficiency requires to sacrifice chemical detail\cite{20},there is ample
evidence\cite{5,20} that a lot of useful insight can be gained already from coarse-grained 
models of flexible polymers such as the bond-fluctuation model\cite{24,25,26}. Each effective
monomer blocks a cube of 8 neighboring sites from further occupancy on a simple cubic lattice.
Effective monomers are connected by effective bond vectors of length $2,\sqrt{5},\sqrt{6},3$ or $\sqrt{10}$
in units of the lattice spacing $a_0\equiv1$. (Each effective bond represents a group of $n\approx3-5$
subsequent $C-C$-bonds along the backbone of the chain.) Using a chain length $N=32$, as 
done here, corresponds to a degree of polymerization of $100-160$ in a real polymer. For a
volume fraction $\Phi=0.5$ of occupied sites, properties of a dense melt are well reproduced\cite{25,26}.
Interactions between monomers are modeled by energy parameters $\epsilon_{AA}=\epsilon_{BB}=-\epsilon_{AB}=-k_BT$,
if the distance between monomers does not exceed $\sqrt{6}$. The statistical segment length $b$
in the relation for the radius of gyration $R_g=b\sqrt{N/6}$ is $b=3.05$ ($R_g\approx 7$ for $N=32$).
The Flory-Huggins parameter is $\chi=2z_{eff}\epsilon$ where $z_{eff}\approx2.65$ denotes the effective coordination number 
in the bulk\cite{17}. The normalized compressibility is $k_BT\kappa a_0^{-3}=4.1 $ \cite{19}.
The bulk phase diagram\cite{25}, the correlation length of concentration fluctuations\cite{27}, etc.\ have been 
determined in previous work.
\newline
Interfaces are studied in a $ L\times D \times L $ geometry where lattice linear dimensions $L$ parallel to 
the interface are $L=512$, and perpendicular $D=64$. Choosing boundary conditions periodic in x,z directions and ``antiperiodic''
in y direction (i.e.\ an A-chain part leaving the box at $ y = -D/2  $ reenters at $ y=+D/2+1 $ as a B-chain, etc.\ ) one 
maintains a system with a single interface at $y=0$.
The system is initialized (for homopolymer interfaces) by choosing chains that
have their center of gravity in the left part of the box ($y<0$) as B-chains, and in the right
part as A-chains. This corresponds to complete segregation between A and B in the bulk,
appropriate for large enough $\epsilon$ (e.g.\ $\epsilon=0.1$). For weakly segregated mixtures, rapid
equilibration of bulk concentrations $\rho_A,\rho_B$ is achieved by use of semi-grandcanonical
algorithms\cite{17,25}. However, very long runs to equilibrate the model system are nevertheless required, 
in order to allow for a build-up of long range capillary  wave fluctuations of the interface. Note, 
that the interfacial position itself
can fluctuate 
and therefore the origin of the y-axis is always
fixed at the center of the (instantaneous) interfacial profile for all ``measurements''.
\newline
For simulating interfaces with adsorbed copolymers, we initialize the system using an equilibrated homopolymer
interface. Then, by choosing 1024 chains (out of 32768) whose center of mass is in the interval $[-\delta,+\delta]$
with $\delta=3$ or $9$, we ``transform'' them into copolymers. The configurations are equilibrated with $2.5 \cdot 10^5$ 
attempted moves per monomer, using a random hopping algorithm\cite{26}. For $\epsilon=0.1$ (where the concentration
of the copolymer in the bulk is $0.04\%$\cite{28}) no effect of varying $\delta$ was found. We average over 86 configurations, 
taken every $10^4$ attempted moves per monomer. We estimate that under the chosen conditions about $30\%$ of the interfacial
area is covered by copolymers, i.e.\ we study the dilute case (``mushroom regime'' rather the ``brush regime''\cite{7}).

\section{Density profiles of monomers and single segments}
\begin{figure}
\vspace*{3cm}
\caption{
Monomer density normalized by the bulk density $\rho_m(y)/\rho_{bulk}$ plotted vs. $y/w_{SSL}$ where 
the (theoretical[21]) interfacial width in the strong segregation limit (SSL) is $w_{SSL}=b/(6\chi)^{1/2}\approx 1.71$
at $\epsilon=0.1$. Shown are the total density $\rho$, A and B monomers separately ($\rho_A,\rho_B$), homopolymers 
irrespective of their nature ($\rho_h$), and A and B monomers belonging to a copolymer ($\rho_{AC},\rho_{BC}$).
The monomer profiles for a pure homopolymer system without copolymers are $\rho^0_A,\rho^0_B$. From Ref.[19].
}
\end{figure}
Fig.\ 1 shows profiles of A and B monomer densities in systems with and without copolymers.
It is seen that in this dilute case A and B monomer profiles are the same, whether copolymers are present or not.
The slight dip of the total density, i.e.\ the enrichment of ``free volume'' in the center of the interface, can be
explained quantitatively in terms of the nonzero
compressibility of the model\cite{17}. Qualitatively, the distribution of the copolymer monomers
agree with experimental results, monomers of type A being concentrated in the A-rich phase, and monomers
of type B in the B-rich phase.
\newline
\begin{figure}
\vspace*{3cm}
\caption{
Homopolymer segment density profiles $\rho_s(y)$ plotted vs. $y/w_{SSL}$, for monomers in the middle of the chain ($\rho_{1/2}$),
at the chain ends ($\rho_e$), and at one and three quarters of the chains ($\rho_{1/4}$) and of all homopolymer monomers ($\rho_h$).
Also shown is the distribution of midpoints between the two ends of the homopolymer $\rho_{ee}$. 
Units are the bulk density $\rho_b$ or $\rho_{sb}=\rho_b/16$ , as indicated. From Ref.[19] .
}
\end{figure}
\begin{figure}
\vspace*{3cm}
\caption{
Block copolymer segment density profiles for densities of A,B monomers in the middle of the chains ($\rho_{1/2A},\rho_{1/2B}$),
at the chain ends ($\rho_{eA},\rho_{eB}$), at one and three quarters of the chain ($\rho_{1/4A},\rho_{1/4B}$), and of all
copolymer monomers ($\rho_{CA},\rho_{CB}$). The inset shows the results of the SCF theory for A monomers. From Ref. [19].
}
\end{figure}
A more detailed picture results from the distribution of single chain segments (Figs.\ 2,3). Chain ends of homopolymers
are enriched at the interface relative to the total homopolymer density profile, as in the case without copolymers\cite{17}.
Profiles of inner segments ($\rho_{1/2}$,$\rho_{1/4}$) are close to the total density profile. However
the ``geometrical midpoint'' distribution $\rho_{ee}$ has a minimum at the interface and a maximum a gyration radius away: 
this tells that typically chains have only one end at the interface and not both ends!
\newline
As expected, copolymer middle segments concentrate at the interface, whereas 
chain ends stretch out in their favorite bulk phase. The distribution of monomers in the middle of a block resembles 
that of the total monomer distribution of the corresponding species of the copolymer. All these features are in 
qualitative agreement with SCF theory (the underestimation of the interfacial width by the SCF theory causes some minor 
qualitative deviations, of course).
\newline
\section{Orientation of chains and of bonds}
\begin{figure}
\vspace*{3cm}
\caption{
Orientational asymmetry parameter $q_e(y)$ of the end-to-end distance plotted vs. $y$ at 6 values of $\epsilon$ as indicated. 
From Ref.[17].
}
\end{figure}
Although homopolymers typically have only one chain end right near the interfacial center,
there is nevertheless a pronounced effect on the orientation of the chains. This is seen
by studying the orientational order parameter (Fig.\ 4)$q_e$ for the end-to-end distance,
$q_e(y)=[3\langle R^2_y\rangle_y - \langle \vec{R}^2\rangle_y ] /2\langle \vec{R}^2\rangle_y$
where the outer index $y$ at the brackets $\langle \cdots \rangle_y$ denotes the center of gravity 
y-coordinate of the considered chain, the inner indices $x,y,z$ denote the Cartesian components $\vec{R}=(R_x,R_y,R_z)$.
One sees that $q_e(y)\approx 0$ (i.e.\ random orientation) in the weak segregation limit (note that
the critical point occurs for $\epsilon_c=0.0143$\cite{25}), while the strongly
negative values of $q_e(y\approx 0)$ in the SSL indicate that parallel components of $\vec{R}$ are 
much larger than perpendicular ones. A similar effect occurs for the gyration radius as well, but there is 
hardly any effect for individual bond vectors\cite{17}. Analyzing the mean square gyration components in the eigensystem
of the gyration tensor, one easily finds that Fig. 4 is not primarily an effect of coil deformation (there occurs only a slight
shrinkage of chain linear dimensions) but due to coil orientation (the instantaneous soap-shaped coils are
oriented with their two longer axis more or less parallel to the interface).
\newline
\begin{figure}
\vspace*{3cm}
\caption{
Mean square end-to-end vector components $\langle R^2\rangle$ with $i=x,y,z$ in units of the average bulk value $b^2N/3$, 
plotted vs. the distance of the center of the end-to-end vector from the interface $y$, at $\epsilon=0.1$. Results are shown for 
homopolymers and copolymers and compared to the SCF prediction for the latter. From Ref.[19].
}
\end{figure}
Copolymers show the inverse behavior, they stretch in the direction perpendicular to the interface (Fig. 5). Both SCF 
and Monte Carlo calculations predict that the effect is strongest for the copolymers centered at about one to two radii 
of gyration away from the interface, and much weaker for those chains located in the wings of the concentrations profile 
or at the middle of the interface. Thus one can picture the copolymer in the latter case as consisting of two almost
independent blocks, which hardly feel the effect of being linked together.
One finds that the vectors connecting
the end of the single A or B blocks are on average hardly oriented\cite{19}.
Only the blocks centered deep in their majority phase ($y/w_{SSL}\approx \pm 7$) stretch perpendicular to the interface, since 
they are pulled toward the interface by the other copolymer end. Also the vector $\vec{D}$ connecting the centers of 
the mass of the A and B blocks is on average oriented and strongly stretched in the negative y direction ($\langle \vec{D}_{y}
\rangle = -3.8, \sqrt{\langle \vec{D}_{y}^2\rangle} = 4.5$).
\newline
Thus in this dilute situation single blocks are mostly not oriented at all: the perpendicular
orientation of whole copolymers results from the arrangement of the two constituent blocks. Diblock copolymers 
resemble dumbbells consisting of two mildly perturbed homopolymer coils, and thus have a conformation similar to that found in the 
copolymer melts in the disordered phase near the order-disorder transition\cite{29}.
\newline
\begin{figure}
\vspace*{3cm}
\caption{
a) Orientational order parameter $q$ for various bonds of copolymers as function of their position $y$. Included are end 
bonds, link bonds (from 16th to 17th monomer), bonds next to the link bonds, and bonds in the middle of a block (from 8th to 9th
and from 24th to 25th monomer)
b) Same as a) but from a SCF calculation. From Ref.[19].
}
\end{figure}
This picture is corroborated by the orientation of individual bonds, 
$q(y)=[3\langle b^2_y\rangle_y - \langle \vec{b}^2\rangle_y ] /2\langle \vec{b}^2\rangle_y$, see Fig. 6. For 
homopolymers $q(y)$ stays extremely small throughout, $q(y\approx 0)\approx -0.01$ to $-0.02$ due to the on average 
parallel orientation of coils close to the interface. One finds a similar behavior for the end bonds or the bonds in the 
block middle for the copolymers, while the link bonds are oriented perpendicular to the interface ($q(y)>0$, in
particular if $y$ is one or two gyration radii away from the interface center). This behavior is reproduced by the SCF theory only 
qualitatively. Note that the SCF profiles for $q(y)$ clearly reflect the two different length scales: the width $w_{SSL}$ 
controls the extent of the central dip, while the gyration radius controls the overall width of the region with nonzero $q$. 
Due to capillary waves fluctuations, this distinction is smeared out in the Monte Carlo results.

\section{Discussion}
As far as corresponding experimental results are available, e.g.\ from the work of Russell and coworkers\cite{14}, experiment
and simulation agree qualitatively. On the other hand, by obtaining simultaneous information on both density profiles of
various monomers along the chains and on bond orientations we clearly can go beyond experiment, resulting in a very clear and
detailed ``picture'' how homopolymer and block copolymer configurations look like. But clearly the present work, restricting
attention to a single chain length (chosen also the same for both homopolymers and copolymers) is a first step only,
and one must already expect a different picture when one studies more concentrated block copolymer layers at interfaces.
Also various asymmetries (different chain length of the blocks, different chain stiffnesses, etc.\ ) deserve attention.

\begin{ack}
We acknowledge support from the Deutsche Forschungsgemeinschaft (DFG)
grants No Bi 314/3 and Bi 314/12, and by the Bundesministerium f\"ur Bildung, 
Wissenschaft, Forschung und Technologie (BMBF grant No 03N8008C). Generous 
access to the CRAY-T3Ds of CINECA (Bologna, Italy) and EPFL (Lausanne, 
Switzerland) is acknowledged. We also thank W. Oed and H. Weber for helpful 
programming advice.
\end{ack}

\newpage
 \setlength{\epsfxsize}{12cm}
 \epsffile{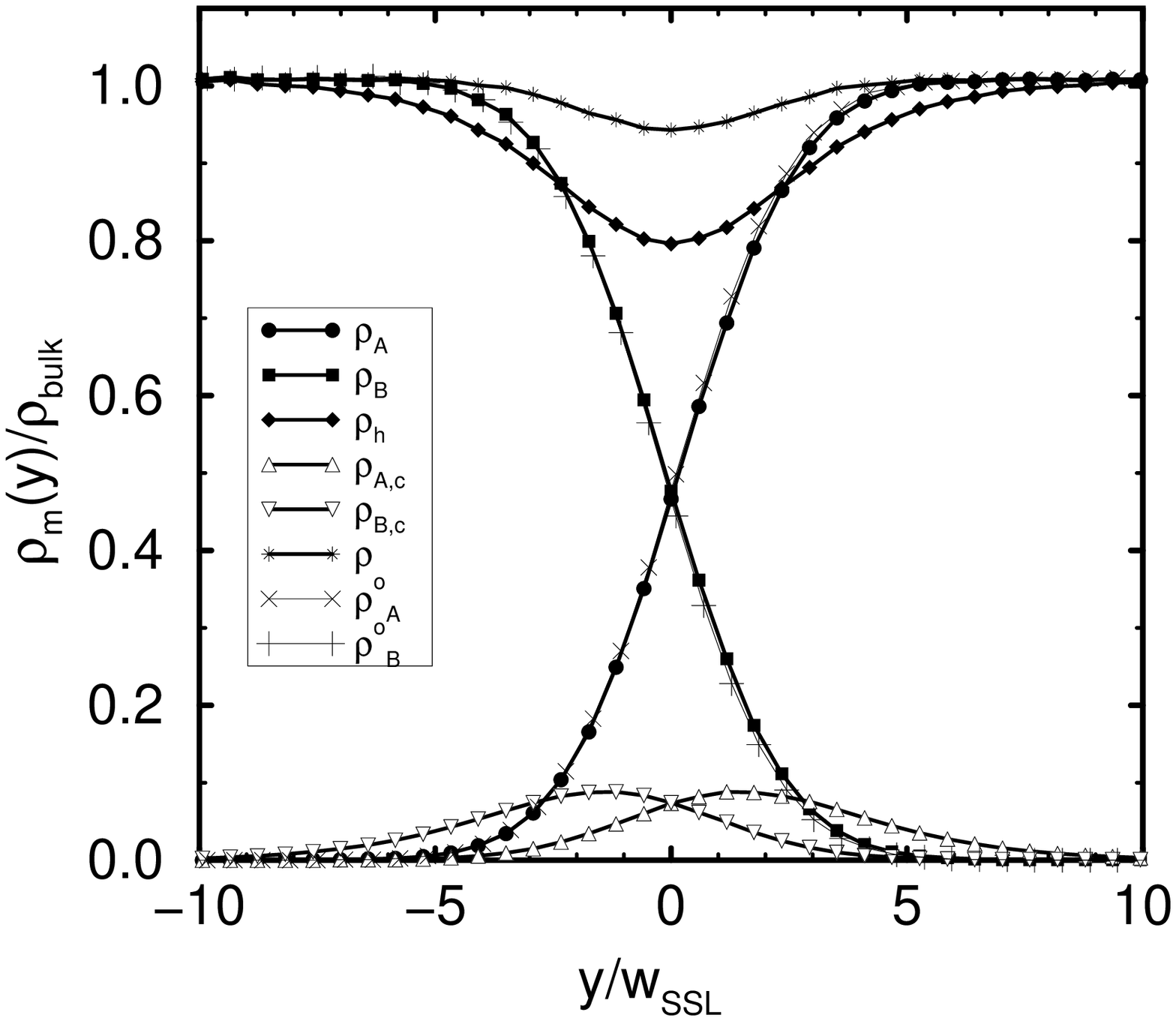}
 Figure 1
\newpage
 \setlength{\epsfxsize}{12cm}
 \epsffile{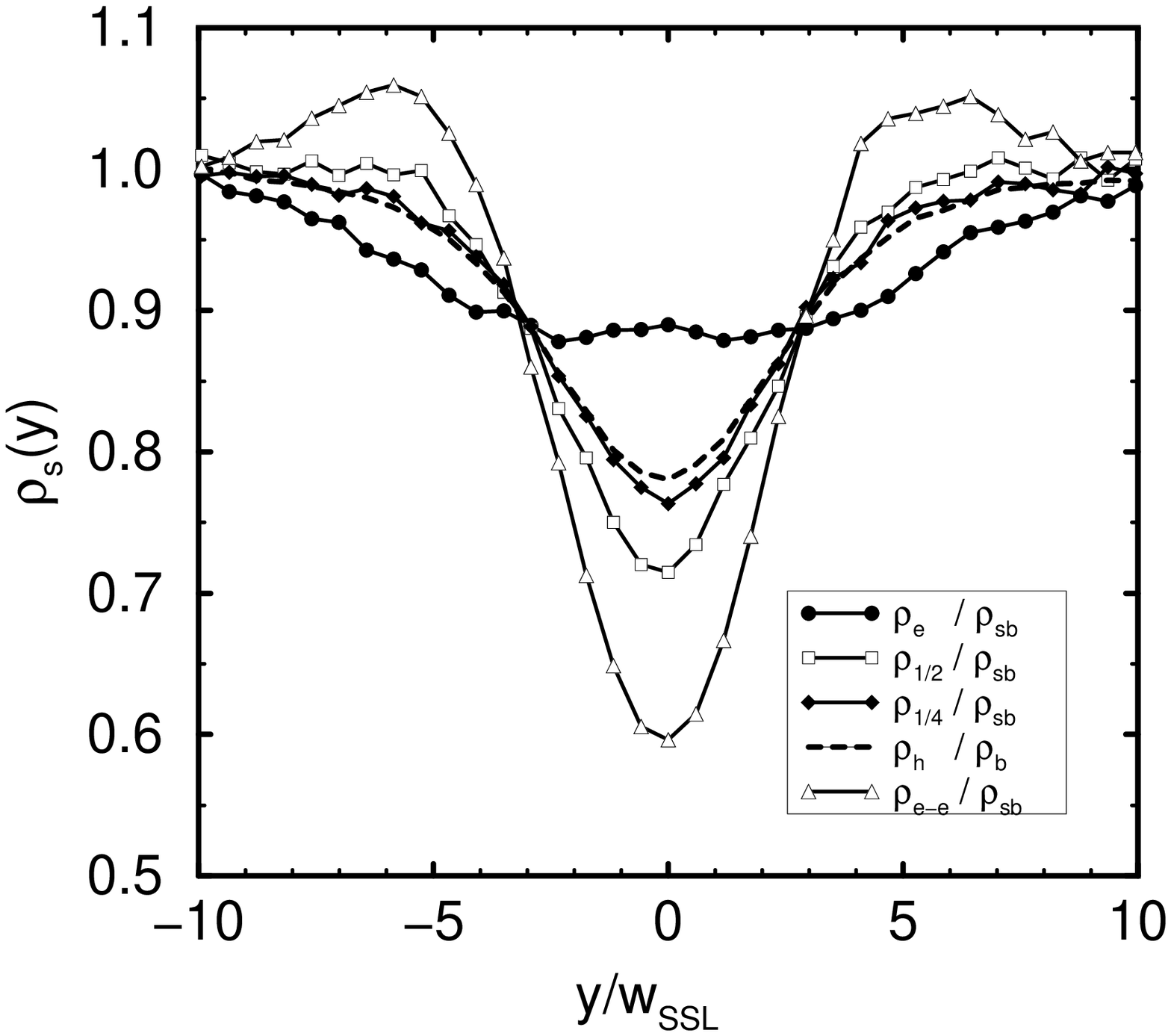}
 Figure 2
\newpage
 \setlength{\epsfxsize}{12cm}
 \epsffile{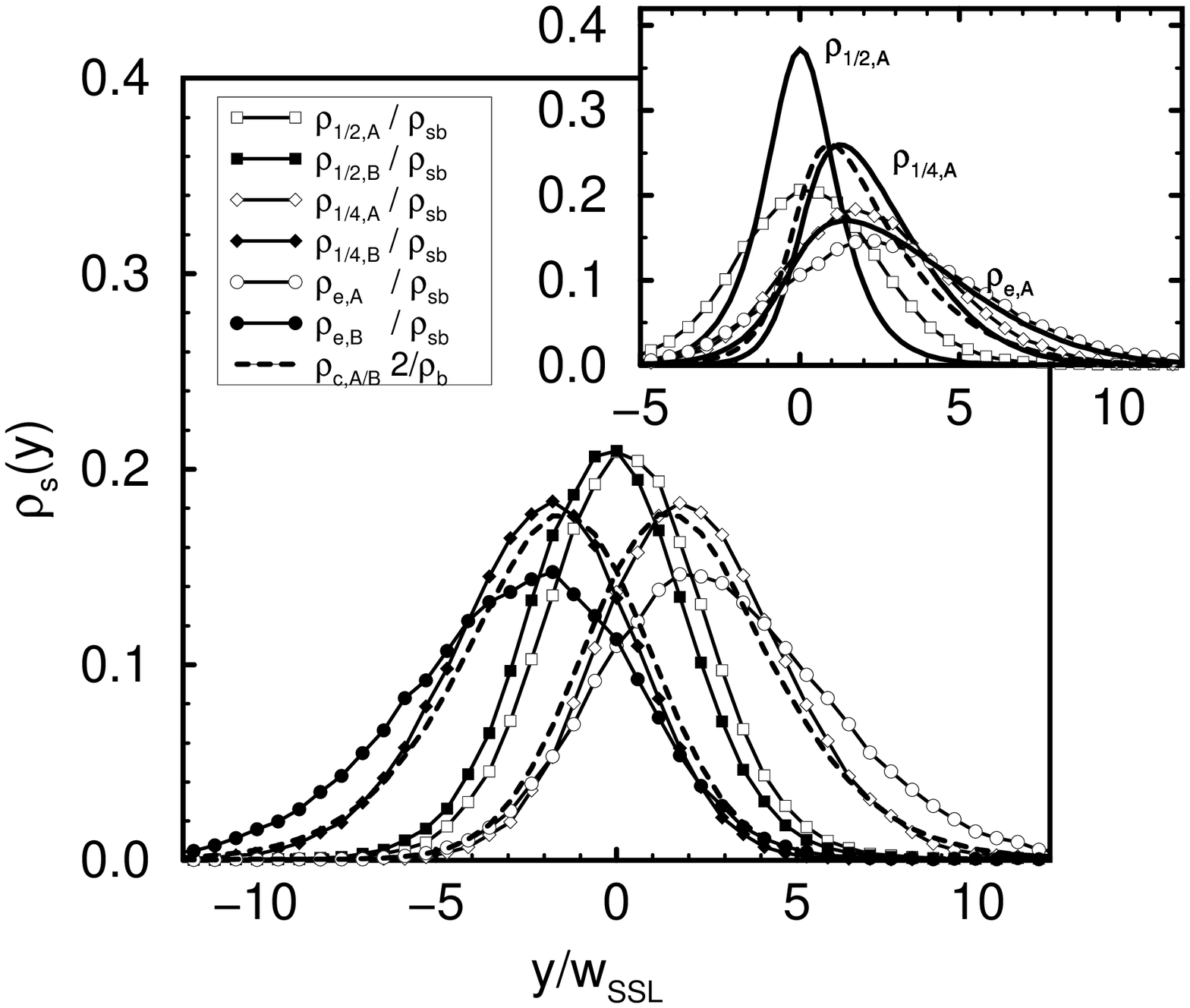}
 Figure 3
\newpage
 \setlength{\epsfxsize}{12cm}
 \epsffile{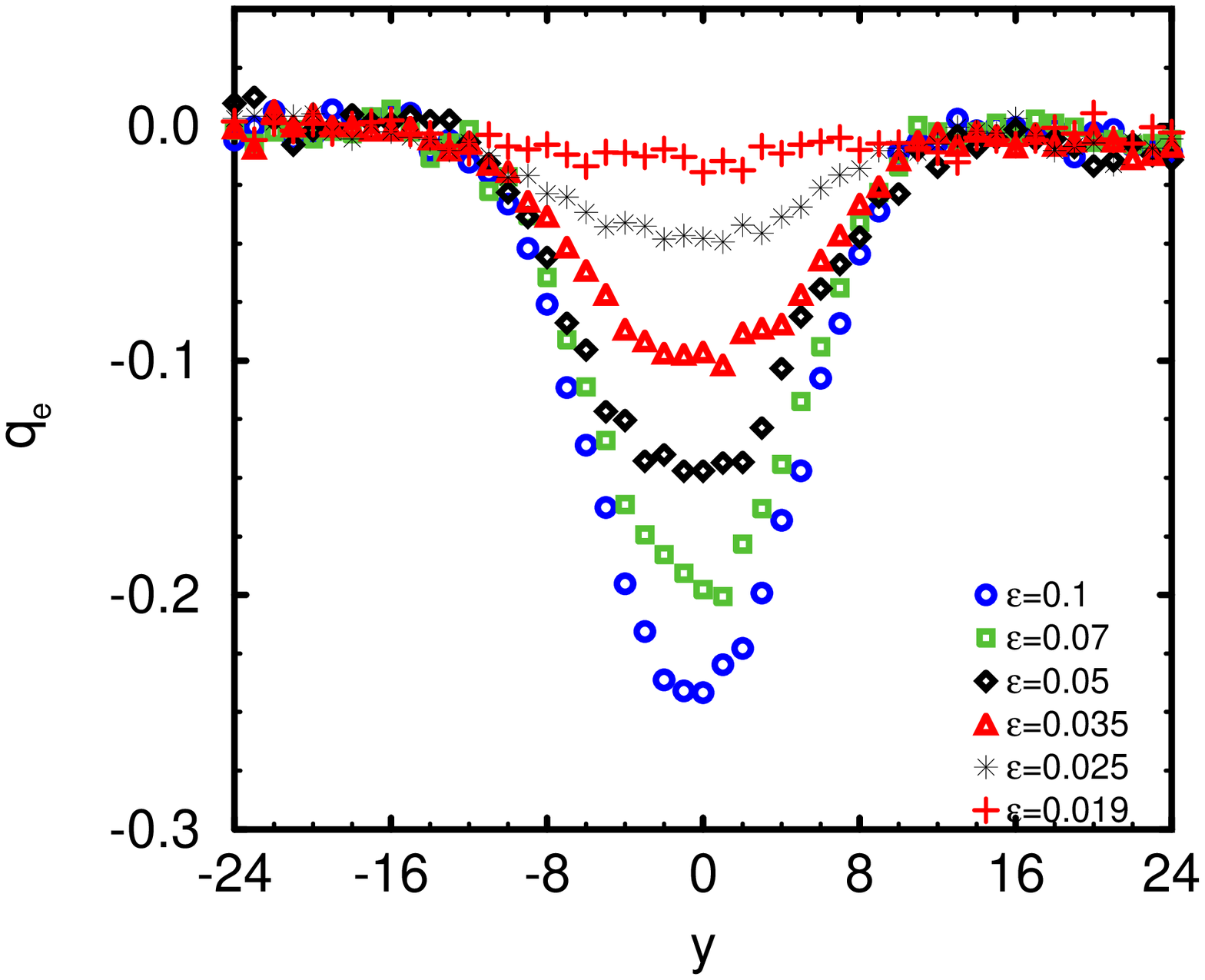}
 Figure 4
\newpage
 \setlength{\epsfxsize}{12cm}
 \epsffile{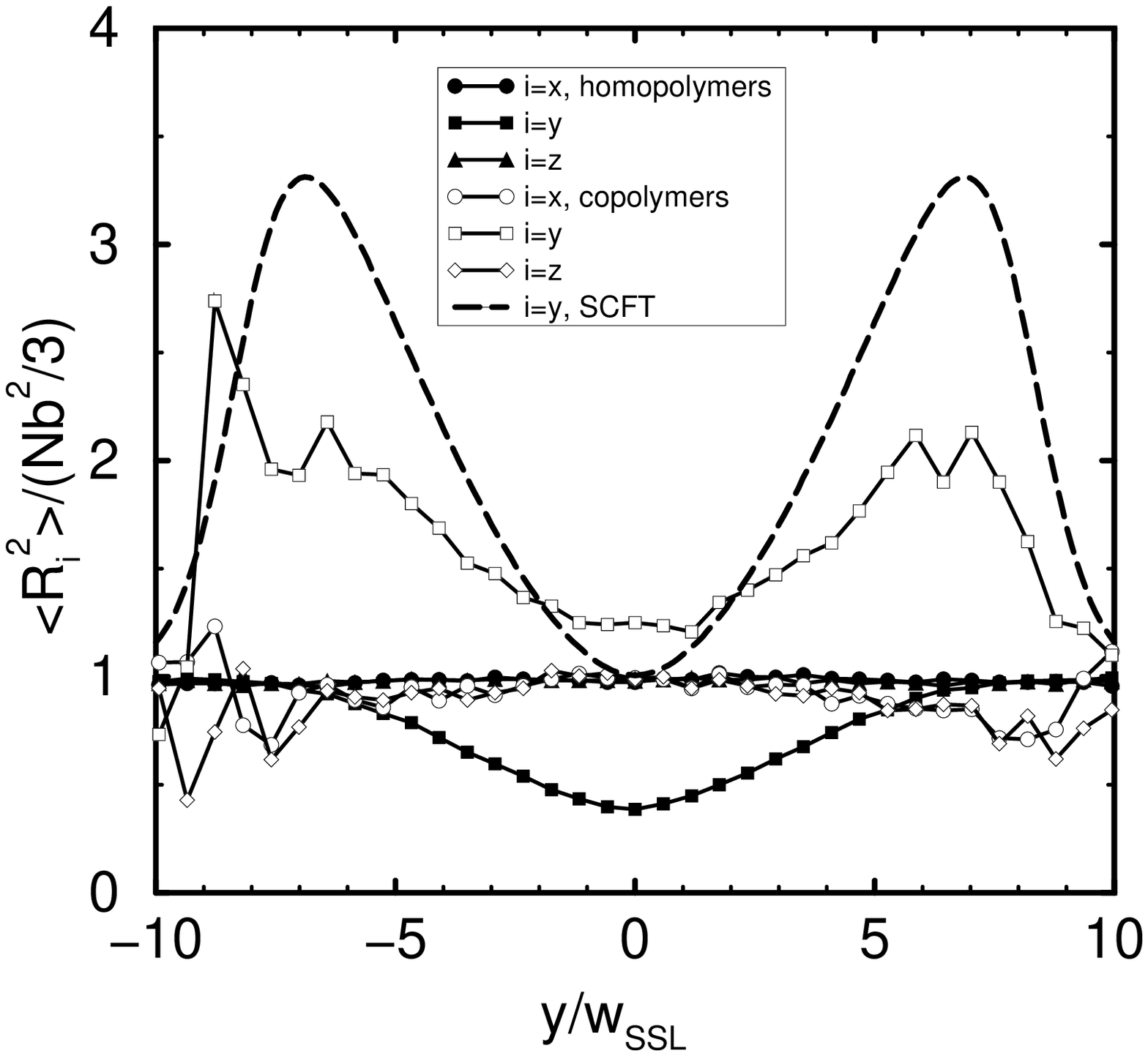}
 Figure 5
\newpage
 \setlength{\epsfxsize}{12cm}
 \epsffile{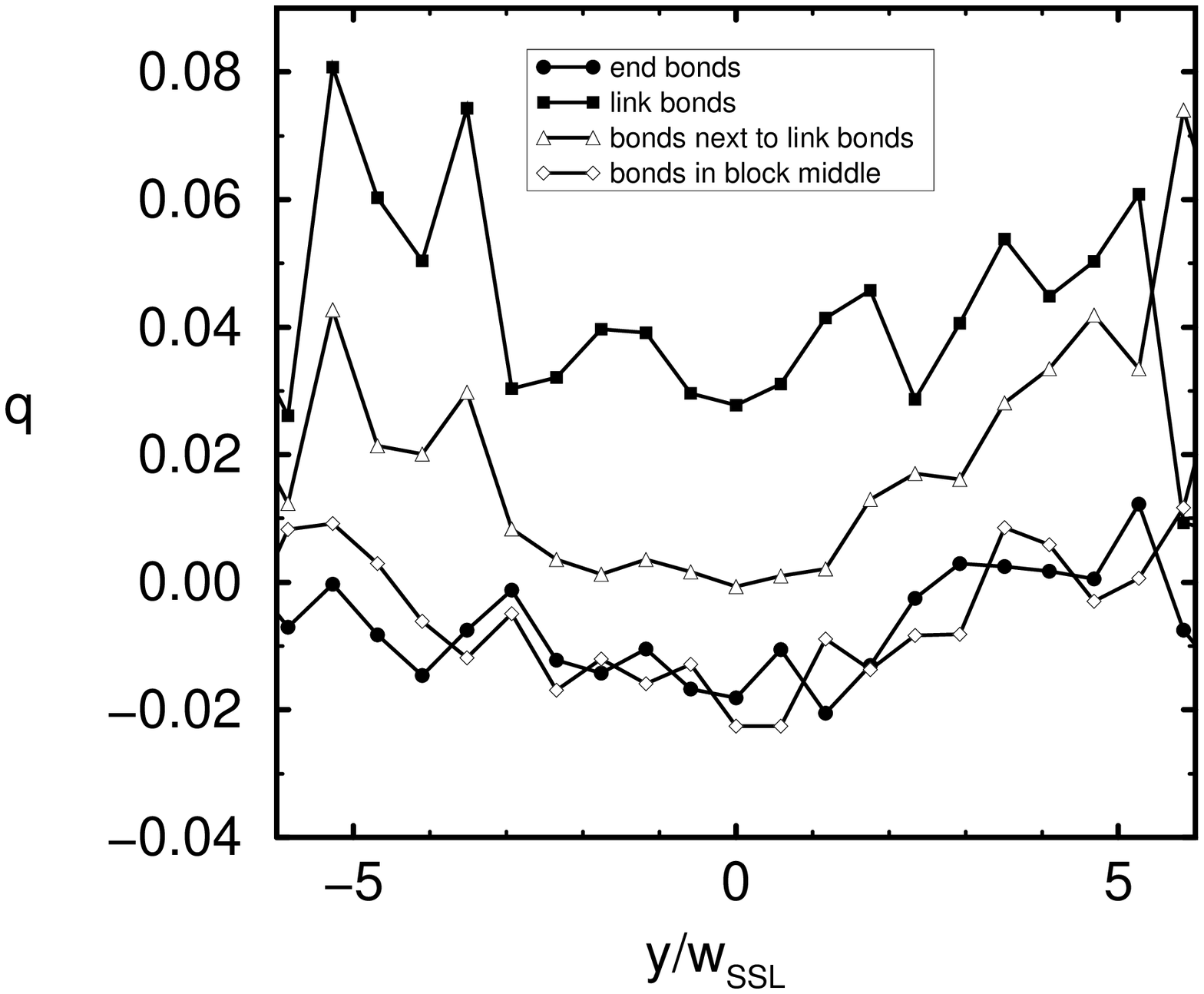}
 Figure 6a

 \setlength{\epsfxsize}{12cm}
 \epsffile{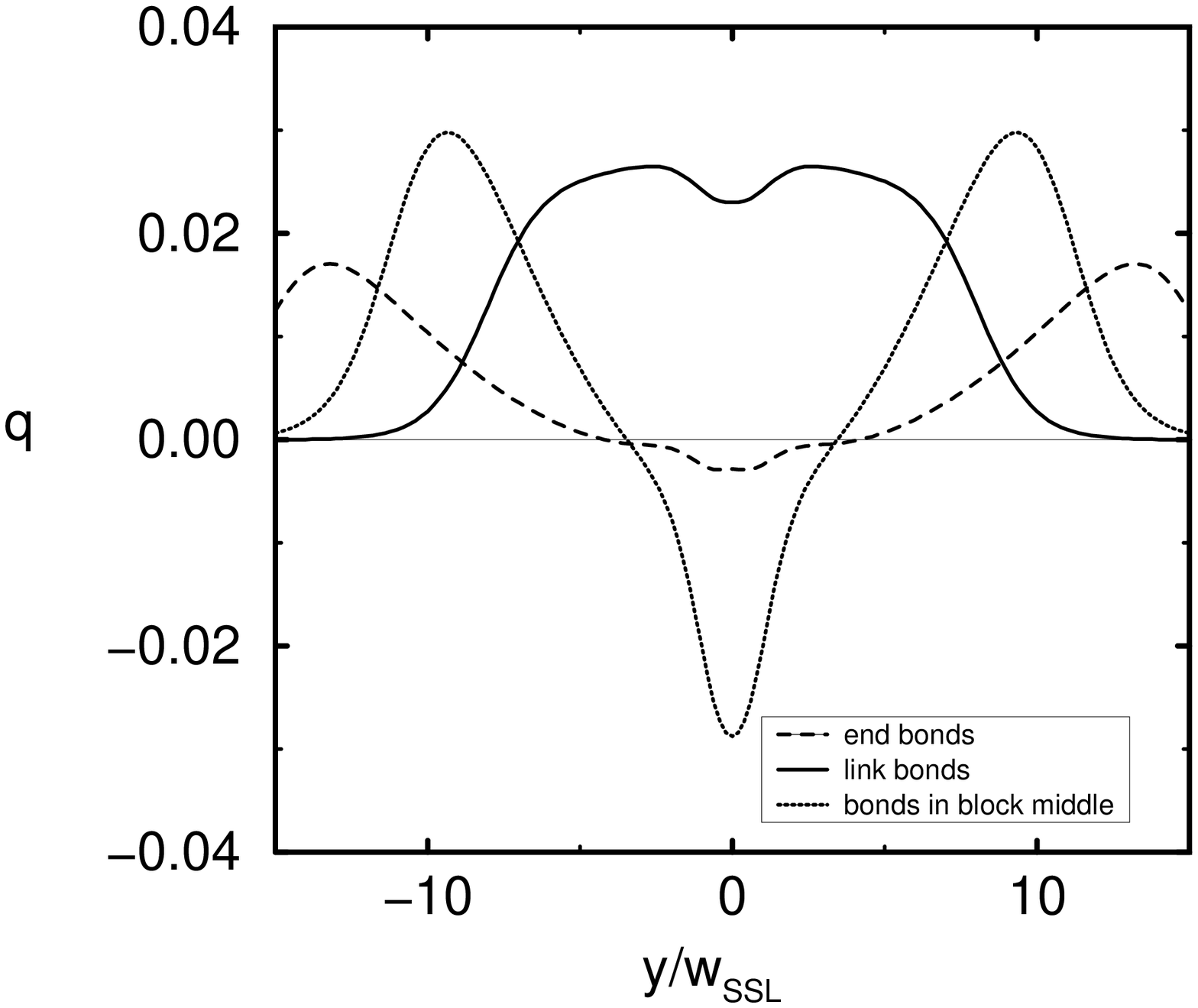}
 Figure 6b
\end{document}